\pdfoutput=1

\documentclass[twocolumn,prb,showpacs,amsmath,amssymb,superscriptaddress]{revtex4}

\bibpunct{[}{]}{,}{n}{}{}

\usepackage{bm}
\usepackage{graphicx}
\usepackage{amssymb}
\usepackage{xcolor}

\newcommand{\beg}{\begin{equation}}
\newcommand{\en}{\end{equation}}

\newcommand{\bq}{\mathbf q}

\begin{document}

\title{Quantum criticality in Ce$_{1-x}$Sm$_x$CoIn$_5$}

\author{D. L. Kunwar }  
\affiliation{Department of Physics, Kent State University, Kent, Ohio, 44242, USA}

\author{R. B. Adhikari}  
\affiliation{Department of Physics, Kent State University, Kent, Ohio, 44242, USA}

\author{N. Pouse}
\affiliation{Center for Advanced Nanoscience, University of California, San Diego, La Jolla, California 92093, USA}

\affiliation{Department of Physics, University of California at San Diego, La Jolla, CA 92903, USA}

\author{M. B. Maple}
\affiliation{Center for Advanced Nanoscience, University of California, San Diego, La Jolla, California 92093, USA}
\affiliation{Department of Physics, University of California at San Diego, La Jolla, CA 92903, USA}

\author{M. Dzero}
\affiliation{Department of Physics, Kent State University, Kent, Ohio, 44242, USA}

\author{C. C. Almasan}
\affiliation{Department of Physics, Kent State University, Kent, Ohio, 44242, USA}

\date{\today}

\begin{abstract}
Motivated by the possibility of observing the co-existence between magnetism and unconventional superconductivity in heavy-fermion Ce$_{1-x}$Sm$_x$CoIn$_5$ alloys, we studied how the samarium substitution on the cerium site affects the magnetic field-tuned-quantum criticality of stoicheometric CeCoIn$_5$ by performing specific heat and resistivity measurements. By applying an external magnetic field, we have observed Fermi-liquid to non-Fermi-liquid crossovers in the temperature dependence of the electronic specific heat normalized by temperature and of the resistivity. We obtained the magnetic-field-induced quantum critical point (QCP) by extrapolating to zero temperature the temperature - magnetic field dependence at which the crossovers take place. Furthermore, a scaling analysis of the electronic specific heat is used to confirm the existence of the QCP. We have found that the magnitude of the magnetic-field-induced QCP decreases with increasing samarium concentration. Our analysis of heat capacity and resistivity data reveals a zero-field QCP for $x_\textrm{cr} \approx 0.15$, which falls inside the region where Sm ions antiferromagnetism and superconductivity co-exist. 
\end{abstract}


\maketitle
\section{Introduction}
CeCoIn$_5$ is an unconventional superconductor in the family of the `115' heavy fermion materials with a fairly high transition critical temperature of T$_c$ at 2.3 K. A consensus exists by now that the unconventional superconductivity in the `115' system is likely governed by its proximity to an antiferromagnetic critical point at zero temperature \cite{petrovic2001heavy,kohori2001nmr}. Generally, these materials can be tuned to a quantum phase transition at a quantum critical point (QCP) by either chemical substitutions \cite{lohneysen1994non,maple1995non}, pressure \cite{mathur1998magnetically,hu2012strong}, or by applying an external magnetic field
\cite{grigera2001magnetic,bianchi2003avoided}.  Consequently, interactions between conduction electrons and critical fluctuations associated with the underlying QCP lead to a manifestation of quite unusual physical properties in both normal and superconducting phases of the `115' materials \cite{coleman2005quantum,stockert1998two}.  The fact that a QCP does exist in these materials is usually elucidated by performing a scaling analysis of the thermodynamic response functions, such as specific heat and magnetic susceptibility. 

A significant amount of experimental data, as well as theoretical results, strongly suggest that the Cooper pairing in CeCoIn$_5$, albeit unconventional, is mediated by the interaction between conduction electrons and localized magnetic moments of partially filled $f$-shells of cerium ions. A tendency towards an antiferromagnetic ordering itself is driven by the exchange interaction between cerium magnetic moments. The exchange interaction - a driver for an antiferromagnetic transition - is also thought to produce soft long-range bosonic modes - a pairing glue for the conduction electrons - to ultimately induce superconductivity with $d$-wave order parameter \cite{van2014direct}. 

Naturally, the chemical substitution of magnetic (and nonmagnetic) rare earth ions for magnetic Ce$^{3+}$ not only allows one to study the effect of inter-site interactions on the QCP, but also separate the single-ion Kondo effect from effects associated with the magnetic exchange interaction. Indeed, the underlying quantum phase transition may shift under the effect of an external magnetic field or pressure. As a result, one may expect that it would affect the superconducting transition temperature, as well as the thermodynamic response functions in the normal state.  Thus, if we were to believe the hypothesis that QCP is governing the unusual transport and thermodynamic properties observed in superconducting and normal states \cite{movshovich2001unconventional,kohori2001nmr,ormeno2002microwave}, systematic studies of the `115'-based alloyed compounds could offer an opportunity to get a deeper insight as well as to unveil differences and/or similarities pertaining to both normal and superconducting states. An additional important aspect of the problem consists in the fact that chemical substitutions inevitably bring disorder into a system, so that the putative QCP may or may not be smeared by the effects associated with the induced spacial inhomogeneities.

Recently, there have been several reports on alloys of Ce$_{1-x}$\emph{M}$_x$CoIn$_5$, where $M$ is a member of lanthanide family with unfilled $f$-orbital shells. The general motivation for studying such alloys lies in an attempt to stabilize the phase of co-existence between superconductivity of the host '115' system and magnetism governed by the impurity magnetic moments. Indeed, one can envision a scenario in which the suppression of superconductivity would be slow enough so that both N\'{e}el ($T_N$) and superconducting ($T_c$) critical temperatures are finite in some region of the phase diagram. In fact, similar effects have been observed in iron-based superconductors \cite{Matsuda_review2014}, although magnetism in those materials is itinerant and the co-existence between $s^{\pm}$-wave superconductivity and spin-density-wave order is possible due to the Anderson theorem \cite{Vavilov2011,Dzero2015}.

Ce$_{1-x}$Yb$_x$CoIn$_5$ is one example of such an alloy: although Yb is supposed to be in the magnetic Yb$^{3+}$ valence configuration, with an increase in ytterbium concentration, its valence configuration
changes from magnetic to a non-magnetic Yb$^{2+}$.
Nevertheless, one of the intriguing results is that the magnetic-field-induced QCP ($H_{\textrm{QCP}}$) of the stoichiometric compound is fully suppressed in the alloy Ce$_{1-x}$Yb$_x$CoIn$_5$ for $x>0.1$ without substantially affecting the unconventional superconductivity and non-Fermi-liquid (NFL) behavior \cite{hu2013non}. Specifically, a zero-field QCP separating paramagnetic and antiferromagnetic phases is observed in Ce$_{0.91}$Yb$_{0.09}$CoIn$_5$ and its presence has been confirmed by the scaling analysis of the specific heat data \cite{singh2018zero}. This is a surprising result given the fact that the dependence of the superconducting critical temperature on ytterbium concentration does not display any correlation with the suppression of the $H_{\textrm{QCP}}$. 

\begin{figure}[h]
\centering
\includegraphics[width=1.0\linewidth]{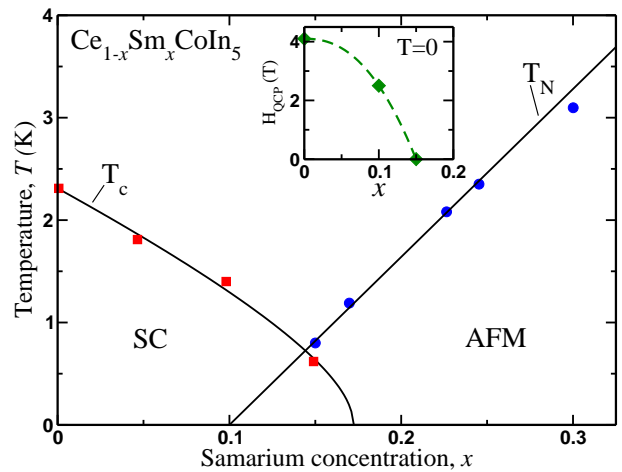}
\caption{(Color online) Temperature - Sm concentration $T-x$ phase diagram of Ce$_{1-x}$Sm$_x$CoIn$_5$. The $T_c(x)$ and $T_N(x)$ data shown here are in part take from Ref. \cite{pouse2018temperature} and from present work. The solid lines are theoretical fits to the experimental data points. We fitted the superconducting critical temperature using Abrikosov-Gor'kov expression for $T_c(x)$ for superconductors with $d$-wave symmetry of the order parameter. The N\'{e}el temperature $T_N$ depends linearly on $x$ at small enough values of $x$ as can be shown by Quantum Monte Carlo simulations and can also be seen by employing a perturbation theory \cite{LinearTN1,LinearTN2}. Inset: Dependence of the magnetic-field-induced quantum critical point $H_{QCP}$, which separates magnetic and non-magnetic states at zero temperature, is shown as a function of samarium concentration. When $x>0.16$ $H_{QCP}$ of stoicheometric CeCoIn$_5$ becomes fully suppressed implying the absence of the anti-ferromagnetic transition in the sub-lattice of Ce moments even at absolute zero temperatures.}
\label{C1F10}
\end{figure}

Another example is provided by Ce$_{1-x}$Sm$_x$CoIn$_5$. Samarium Sm$^{3+}$ replaces Ce$^{3+}$ in CeCoIn$_5$ and, unlike Yb, it remains in the same integer valence configuration. One needs to keep in mind that existing `electron-hole symmetry' between 4$f^1$ (Ce$^{3+})$ and 4$f^{5} ($Sm$^{3+})$ valence configurations implies that larger magnetic moments are introduced into the system through this substitution without changing the carrier density. The phase diagram in the temperature-doping $(T-x)$ plane can be generated based on the $T_N$ and $T_c$ measured from resistivity $\rho(x)$ and  specific heat $C/T$ of  Ce$_{1-x}$Sm$_x$CoIn$_5$ for $0\leq x \leq 0.3$ in zero field, Fig. \ref{C1F10}. 
Therefore, in contrast to Ce$_{1-x}$Yb$_x$CoIn$_5$,  superconductivity in this alloy is completely suppressed at $x\approx 0.18$ and long-range antiferromagnetic (AFM) order emerges in the sub-lattice of Sm moments in Ce$_{1-x}$Sm$_x$CoIn$_5$ for $x\approx 0.10$. While fairly fast (in comparison with Ce$_{1-x}$Yb$_x$CoIn$_5$ alloys), the suppression of superconductivity is not surprising given the sizable magnetic moment of samarium. The fact that the AFM order emerges before superconductivity has been fully suppressed does provide long thought playground to investigate possible co-existence between unconventional superconductivity and magnetism. Furthermore, at the critical concentration $x_{\textrm{cr}}^{\textrm{Sm}}\approx0.1$ where the N\'{e}el temperature for the samarium sublattice vanishes, an antiferromagnetic QCP may be also present. 

Samarium substitutions must also affect the magnitude of $H_{\textrm{QCP}}$ separating the anti-ferromagnetic and paramagnetic phases of cerium sub-lattice at zero temperature. Specifically, as in the case of Yb substitutions, one may expect that the value of $H_{\textrm{QCP}}$ will be suppressed to zero for some critical concentration $x_{\textrm{cr}}$ of samarium ions,
$H_{\textrm{QCP}}(x_{\textrm{cr}})=0$. In the inset of Fig. 1 we see that the magnetic field induced quantum phase transition of the cerium lattice is clearly affected by the samarium substitutions.  In the light of our discussion above, one may then wonder whether $x_{\textrm{cr}}$ satisfies $x_{\textrm{cr}}^{\textrm{Sm}}\leq x_{\textrm{cr}}\leq x^*$ (with $T_c(x^*)=0$) falls into the region where antiferromagnetism induced by the ordering of Sm moments and superconductivity may co-exist or not.  These considerations have lead us to select the  samples $x = 0.10$ and $x = 0.15$ to investigate the changes in the magnitude of $H_{\textrm{QCP}}$ through specific heat and resistivity measurements done in the presence of an applied magnetic field ($H$). In what follows, we present the results of specific heat and resistivity measurements on Ce$_{0.9}$Sm$_{0.10}$CoIn${_5}$ and Ce$_{0.85}$Sm$_{0.15}$CoIn${_5}$ to elucidate the role played by quantum critical fluctuations and their effect on unconventional superconductivity. Our investigation reveals that quantum critical fluctuations strongly correlate with unconventional superconductivity. The field driven QCP is suppressed with increasing Sm concentration. The observed pronounced  crossovers from FL to NFL behavior, as well as the scaling analysis of the $C_e/T$ on both alloys show that the normal state of Ce$_{0.85}$Sm$_{0.15}$CoIn$_5$ is quantum critical. 
 
\section{Experimental Details}
Single crystals of Ce$_{1-x}$Sm$_x$CoIn$_5$ ($x= 0.10$ and $0.15$) were grown using the molten In flux method in alumina crucibles, as described in Ref. \cite{zapf2001coexistence}. The composition and crystal structure were determined from X-ray powder diffraction (XRD) and energy dispersive X-ray (EDX) techniques. The crystals' actual chemical composition is the same as that of the nominal doping, as confirmed by the EDX using FFI Quanta 600 scanning electron microscope equipped with an INCA EDX detector from Oxford Instruments \cite{pouse2018temperature}. 

The crystals were cut into a typical size of 2.0x0.5x0.17 mm$^3$,  with  the $c$-axis  along  the  shortest dimension of the crystals. These single crystals were first polished  with  sandpaper  and  then  etched  in  a  5\%  HCl solution  for  three  hours  to  remove  the  indium  left  on the surface during the growth process.  Then they were washed thoroughly with ethanol to remove any acid and impurities.

Heat capacity measurement was performed in an external magnetic field in the range $0\leq H\leq 14$ T applied perpendicular to the $ab$-plane of each crystal and in the temperature range 0.42 K $\leq T\leq 8$ K. The data were obtained using a relaxation technique in the He-3 option of the Quantum Design Physical Properties Measurement System.  

Four gold leads were attached to each crystal using silver-based epoxy, with the current $I \parallel a$-axis of the single crystal. We performed temperature ($T$) dependent electrical resistivity $(\rho(T))$ measurements between 0.5 K and 300 K in zero magnetic field to extract the superconducting  transition temperature and the Kondo lattice coherence peaks. The resistivity in field  $(\rho(H))$ was measured by scanning temperature from  0.6 K  to 10 K for selected field values from 6 T to 14 T. Then in-plane transverse magnetoresistivity ($\Delta \rho_{a}^{\perp}/\rho_a$)  was measured by scanning the field from -14 T to 14 T for selected temperatures between 2 and 50 K.  
\section{Results}
\subsection{Ce$_{0.90}$Sm$_{0.10}$CoIn$_5$}
\begin{figure}
\centering
\includegraphics[width=1.0\linewidth]{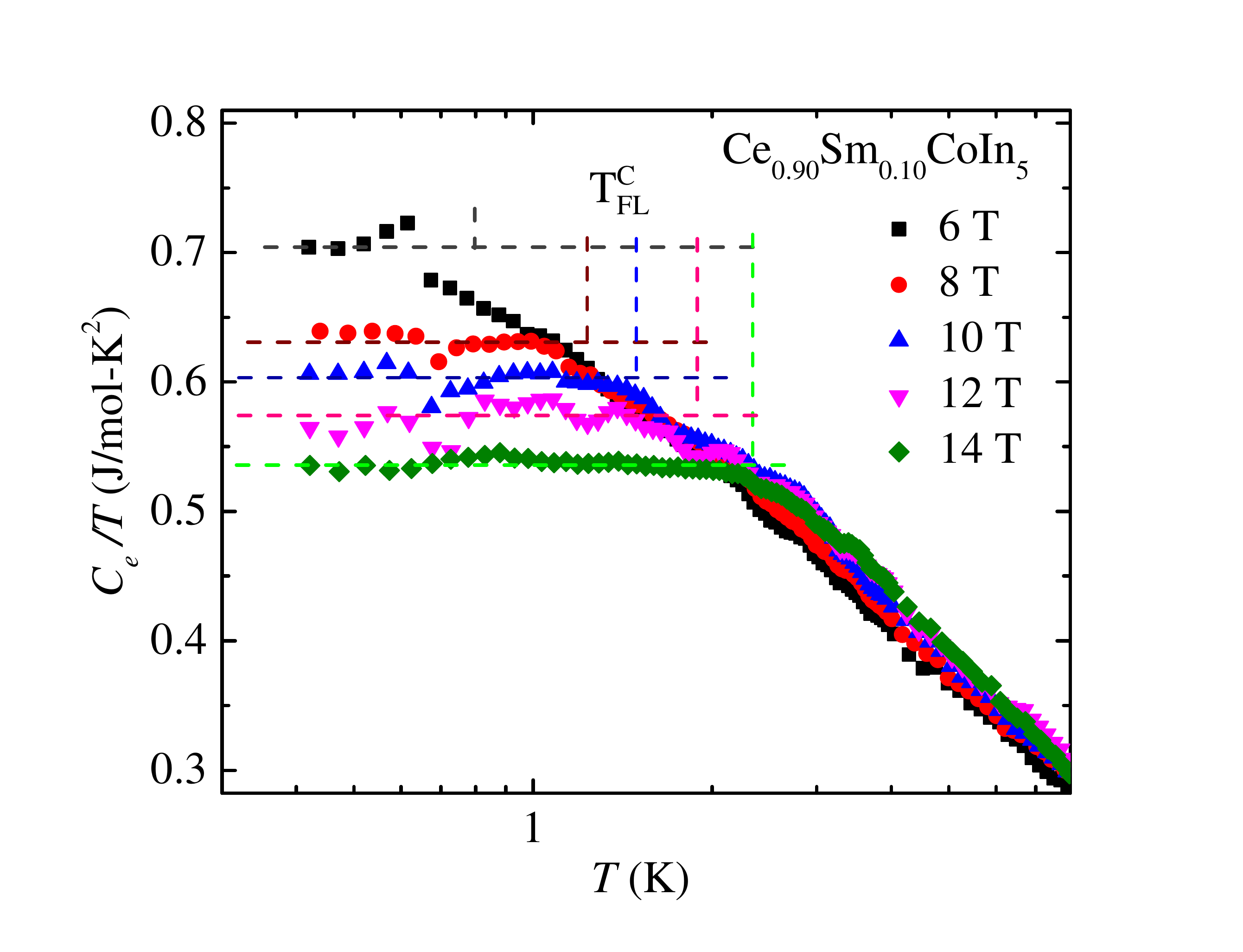}
\caption{(Color online) Semi-Log plot of  the electronic specific heat normalized by temperature $T$,
$C_e/T$  vs $T$ of Ce$_{0.90}$Sm$_{0.10}$CoIn${_5}$ measured in applied magnetic field $H\parallel c$ axis  over the temperature range 0.42 K $\leq T\leq 10$ K. $T^C_{FL}$ represents the temperature at which the behavior changes from  Fermi-liquid to non-Fermi liquid.}
\label{C1F1}
\end{figure}

\begin{figure}
\centering
\includegraphics[width=1\linewidth]{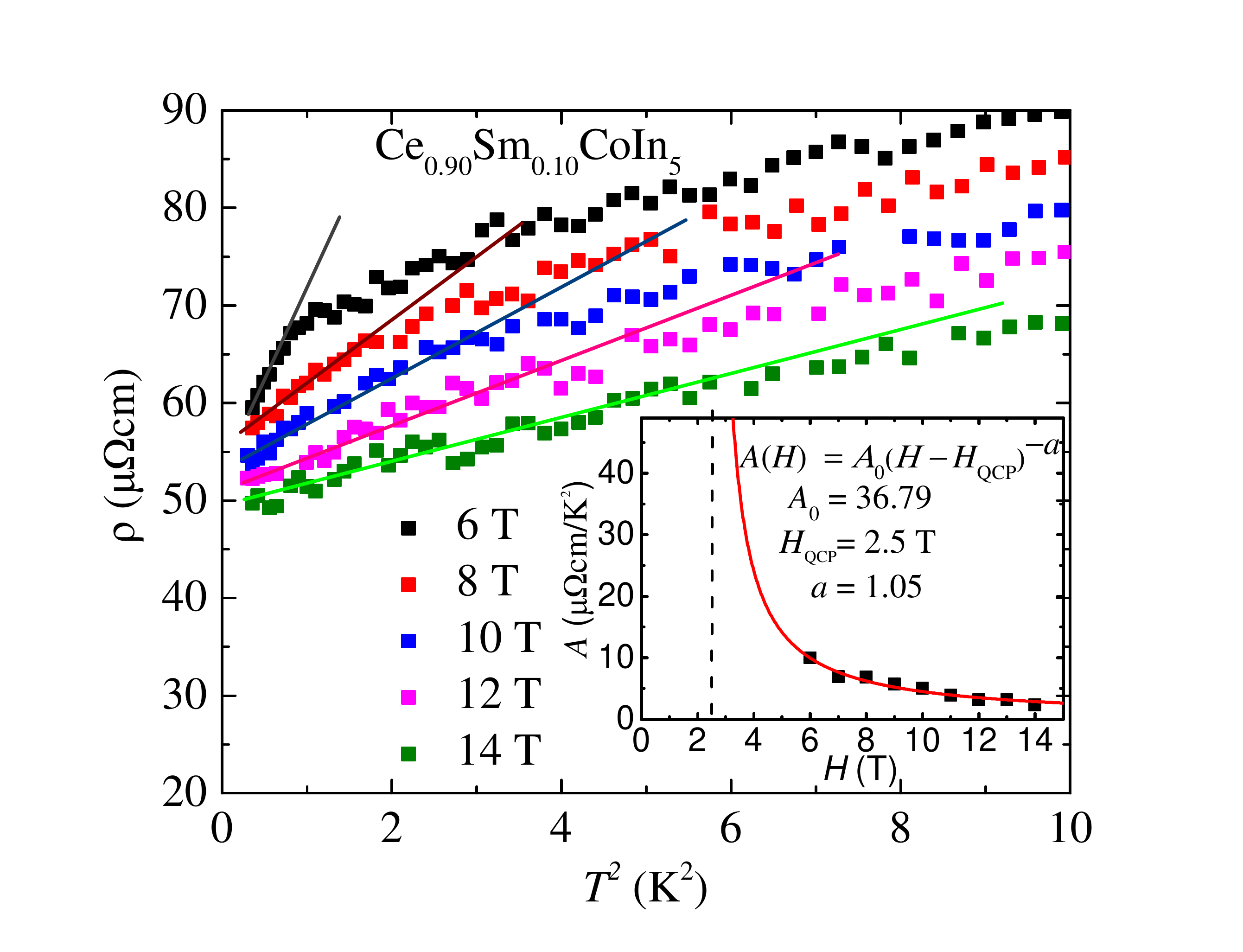}
\caption{ (Color online)  Low temperature $T$ resistivity $\rho$ plotted as a function of T$^2$ in the field range 6 T$\leq H\leq$ 14 T for Ce$_{0.90}$Sm$_{0.10}$CoIn${_5}$. The solid lines represent linear fits  of the data with $\rho(H) = \rho_{0} + AT{^2}$. Inset: Plot of the slope $A$ as a function of magnetic field $H$. The red solid line is a fit of the data.}
\label{C1F2}
\end{figure}

\paragraph{Specific heat data.} Figure~\ref{C1F1} shows the temperature dependence of the electronic specific heat normalized by temperature $C_e(T)/T$ of Ce$_{0.90}$Sm$_{0.10}$CoIn$_5$ measured over the field range 6 T $\leq H\leq$ 14 T and temperature range  0.4 K $\leq T\leq$ 8 K. We have subtracted the specific heat \cite{pouse2018temperature} of the non-magnetic reference compound LaCoIn$_5$ from the measured specific heat to get the electronic and magnetic contributions to the specific heat. Then, to obtain the electronic contribution to the specific heat, we subtracted the high temperature tail of the Schottky anomaly due to quadrupolar and magnetic spin splitting of Co and In nuclei \cite{movshovich2001unconventional}.  All the data shown in Fig.~\ref{C1F1} are normal state results since superconductivity is suppressed in this temperature range measured at fields above 4 T (see purple data taken at 4 T in the inset to Fig. 6). 

Applying an external magnetic field, we observed definite crossovers from constant $C_e/T$ vs $T$ at low temperatures, i.e. Fermi-liquid behavior, to logarithmic temperature-dependent $C_e/T$ with $C_e/T\propto$ ln T$^{-0.4}$ at high temperatures, i.e., non-Fermi liquid behavior. For the measured temperature range, the clearly visible FL regime ($C_e/T$ = constant) at low temperatures is observed for $H \geq 6$ T. The data taken in all these applied magnetic fields exhibit the FL to  NFL crossovers at a temperature $T^{\mathrm{C}}_{\mathrm{FL}}$ that shifts towards higher temperatures with increasing $H$, as shown by the vertical dashed lines of Fig.~\ref{C1F1}. The $T^{\mathrm{C}}_{\mathrm{FL}}$ from each specific heat data measured at different fields are extracted and plotted as shown in the $H-T$ phase diagram of Fig.~\ref{C1F4}.

\paragraph{Resistivity data.} Figure~\ref{C1F2} shows the $T^2$ dependence of resistivity of Ce$_{0.90}$Sm$_{0.10}$CoIn$_5$ measured in 
different $H$. The resistivity is linear in $T^2$ at low temperatures: $\rho(H,T) = \rho_{0} + AT^2$, where $\rho_{0}$ is the residual resistivity, and $A(H)$ is a constant that measures the strength of electron-electron interactions. The linear in $T^2$ behavior of resistivity at low $T$ and high $H$ reveals the recovery of the FL behavior. Its deviation from linearity with increasing temperature is the signature of the NFL behavior \cite{paglione2003field}. The crossovers from linear (FL) to non-linear (NFL) $T^2$ dependence of $\rho(H,T)$  at low temperatures represented by $T^{\mathrm{R}}_{\mathrm{FL}}$, shift towards higher temperatures with increasing magnetic field. The $T^{\mathrm{R}}_{\mathrm{FL}}$ from each resistivity data measured at different fields are extracted and plotted as shown in the $H-T$ phase diagram of Fig.~\ref{C1F4}.

The slope $A$ of the linear fit (solid lines in Fig.~\ref{C1F2}) increases with decreasing field (see inset to Fig.~\ref{C1F2}). Its field dependence follows $A(H) \propto(H-H_{\textrm{QCP}})^{-a}$ with $H_{\textrm{QCP}}$ = 2.5 T and  $a$ =1.05. Thus, this fitting reveals the quantum critical behavior for this Sm doping, with a field-induced quantum critical point of 2.5 T. We note that an $H_{\textrm{QCP}} \approx 5$ T  has been extracted through a similar type of analysis of resistivity data in the parent compound CeCoIn$_5$ \cite{paglione2003field,bianchi2003avoided}. As we will show below, an extrapolation to zero temperature of the linear fit of $T^{\mathrm{C}}_{\mathrm{FL}}(H,T)$ and $T^{\mathrm{R}}_{\mathrm{FL}}(H,T)$ vs $T$ also gives $H_{QCP}= 2.5$ T (see Fig.~\ref{C1F4}).

\begin{figure}
\begin{center}
\includegraphics[width=1\linewidth]{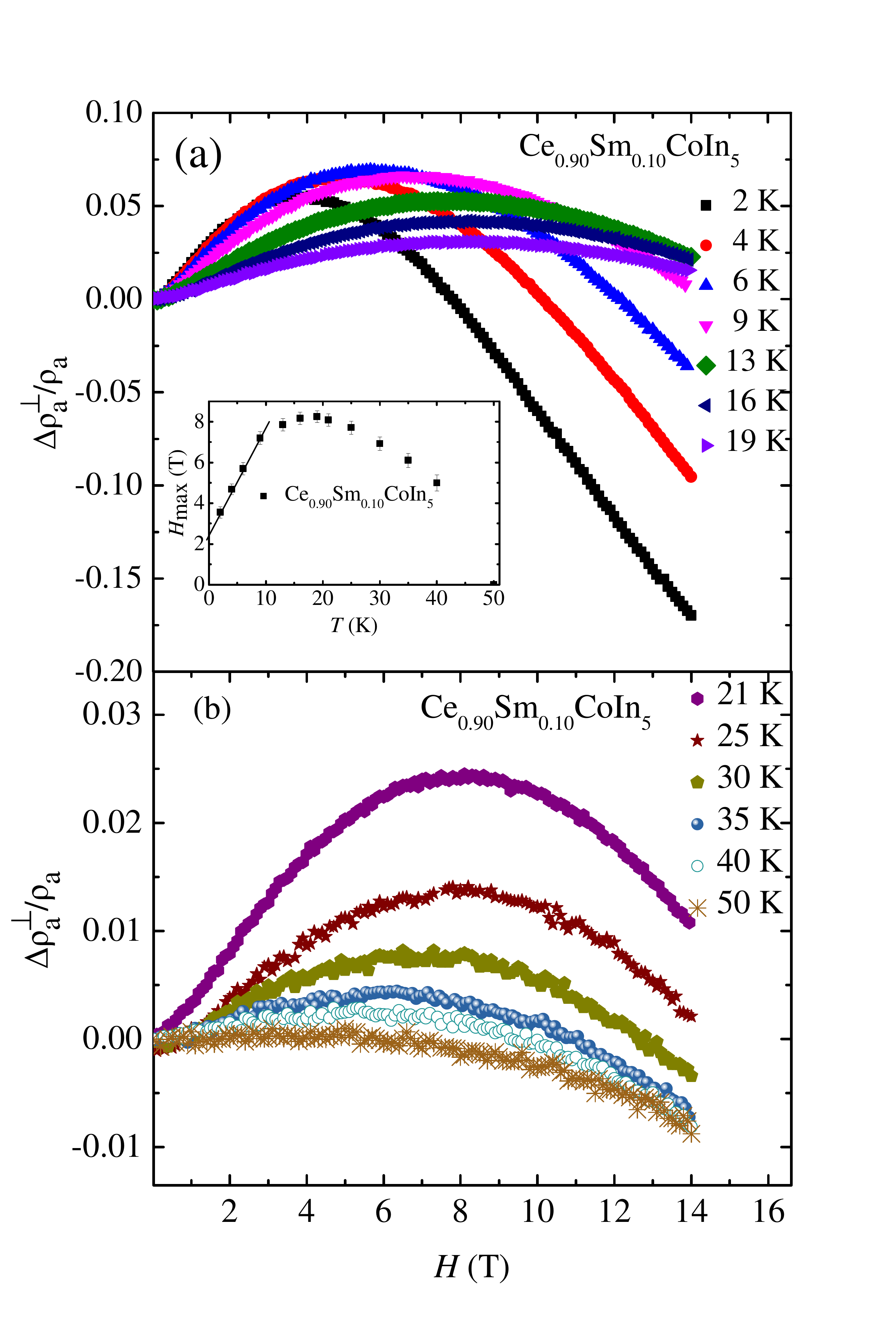}
\caption{(Color online) (a) In-plane transverse  magnetoresistivity $\Delta \rho_a^{\perp}/\rho_a$ vs magnetic field $H$ of Ce$_{0.90}$Sm$_{0.10}$CoIn$_5$ measured over a temperature $T$ range 2 K$\leq T\leq$ 19 K. Inset: $T$ dependence of the characteristic field $H_{max}$ corresponding to the maximum of the transverse MR. The solid line in the inset is the linear fit of the low $T$ data. (b) $\Delta \rho_{a}^{\perp}/\rho_a$ vs $H$ measured over a $T$ range 21 K$\leq T\leq$ 50 K}
\label{C1F3}
\end{center}
\end{figure}

\begin{figure}
\begin{center}
\includegraphics[width=3.3in,angle=0]{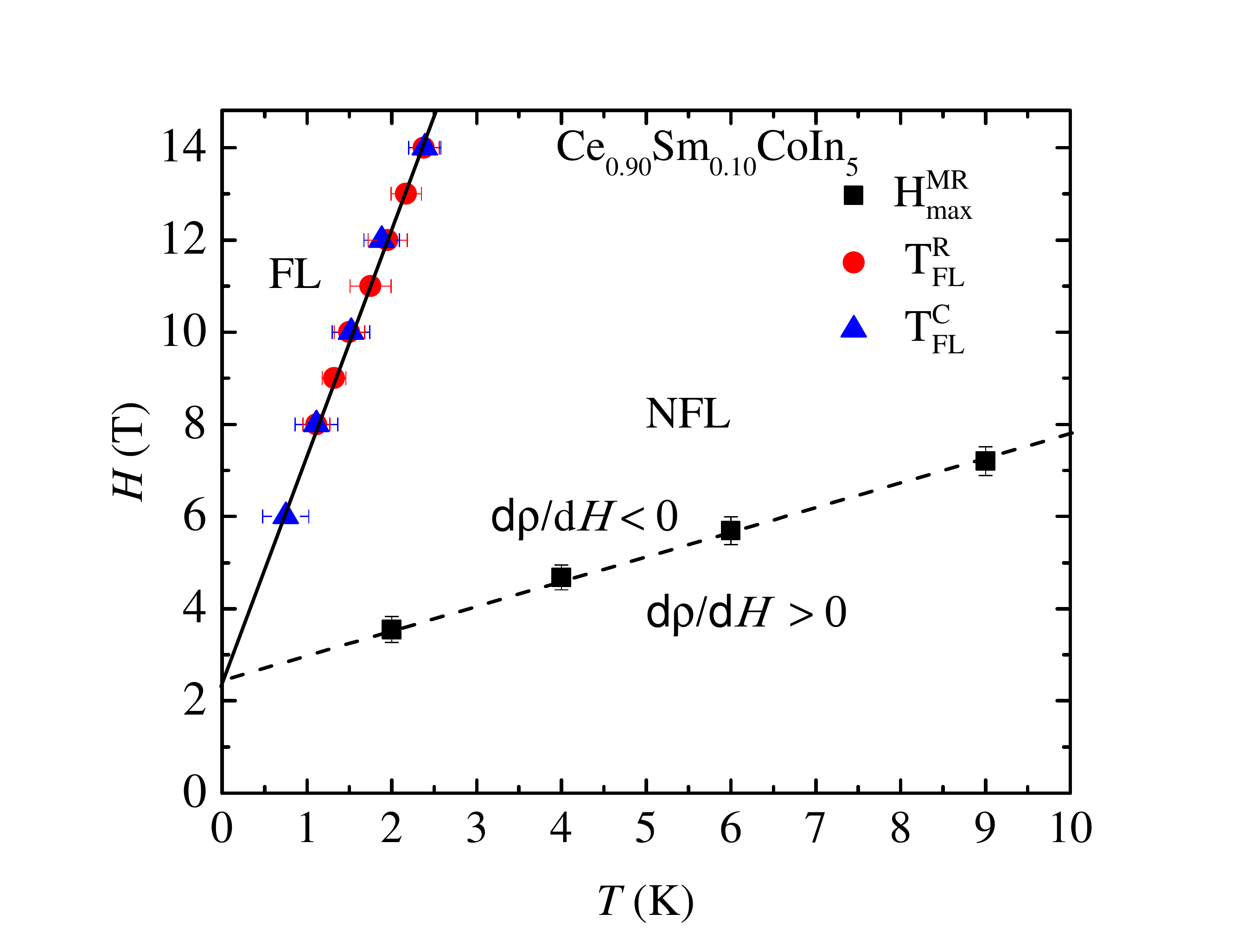}
\caption{(Color online) Magnetic field-temperature $(H-T)$ phase diagram of Ce$_{0.90}$Sm$_{0.10}$CoIn$_5$ with $H\perp ab$ plane. The solid and dash lines are linear fits of the the data extrapolated to zero temperature.}
\label{C1F4}
\end{center}
\end{figure}

\paragraph{Magnetoresistivity data.} Figures~\ref{C1F3}(a) and~\ref{C1F3}(b)  show the in-plane transverse  magnetoresistivity (MR) $\Delta\rho_{a}^{\perp}/\rho_a$ vs $H$ of  Ce$_{0.90}$Sm$_{0.10}$CoIn$_5$ measured over the temperature range 2 K $\leq T\leq$ 50 K. These figures reveal pronounced crossovers from positive to negative MR that become flatter with increasing temperature. The peaks in MR take place at temperatures smaller than the coherence temperature $T_{\textrm{coh}}=41$ K. 

We extracted the  field $H_{\textrm{max}}$ at which the magnetoresistivity is maximum [see main panel of  Fig.~\ref{C1F3}(a) and ~\ref{C1F3}(b)] and plotted $H_{\textrm{max}}$ vs $T$  in the inset to Fig.~\ref{C1F3}(a). This plot shows a non-monotonic dependence of  $H_{\textrm{max}}$ vs $T$, with a maximum at around 19 K. For $T>19$ K, the value of $H_{\textrm{max}}$ decreases with increasing temperature, revealing the coherent Kondo lattice behavior in this $T$ range. However, the positive MR that decreases with decreasing $T$ in the low temperature range ($T<$ 19 K) indicates unconventional Kondo lattice behavior and is attributed to field quenching of the AFM spin fluctuations \cite{malinowski2005c} responsible for the NFL behavior, previously observed in CeCoIn$_{5}$ \cite{malinowski2005c} and Ce$_{1-x}$Yb$_x$CoIn$_{5}$ \cite{hu2013non}. Therefore, the positive MR measured for the $x = 0.10$ single crystals at $T<$ 19 K is due to the AFM spin fluctuations. The extrapolation of the low $T$ linear fit of $H_{\textrm{max}}$ vs $T$ to zero temperature give $H_{QCP}=2.5$ T for this Sm doping.

\paragraph{Magnetic-field-induced quantum critical point.} Based on features extracted from $C_e/T$, $\rho$, and $\Delta \rho_a^{\perp}/\rho_a$ data, we generated the $H-T$ phase diagram shown in Fig.~\ref{C1F4}. Specifically, the FL to NFL  crossover temperatures for different $H$ values are extracted from the $C_e/T$ vs $T$ data of Fig.~ \ref{C1F1} as $T^{\mathrm{C}}_{\mathrm{FL}}$ (blue triangles) and $\rho$ vs $T^2$ data of Fig.~\ref{C1F2} as $T^{\mathrm{R}}_{\mathrm{FL}}$ (red circles). The linear fit of $T^{\mathrm{C}}_{\mathrm{FL}}$ and $T^{\mathrm{R}}_{\mathrm{FL}}$ from both measurements extrapolates to zero temperature at $H_{\textrm{QCP}}=2.5$ T. In addition, the linear extrapolation of the peak  $H^{\mathrm{MR}}_{\mathrm{max}}$ (black squares) vs $T$ (extracted from $\Delta \rho_a^{\perp}/\rho_a$ of Fig.~\ref{C1F3}) to zero temperature gives the same $H_{\textrm{QCP}} =2.5$ T.

\begin{figure}[!]
\includegraphics[width=3.3in,angle=0]{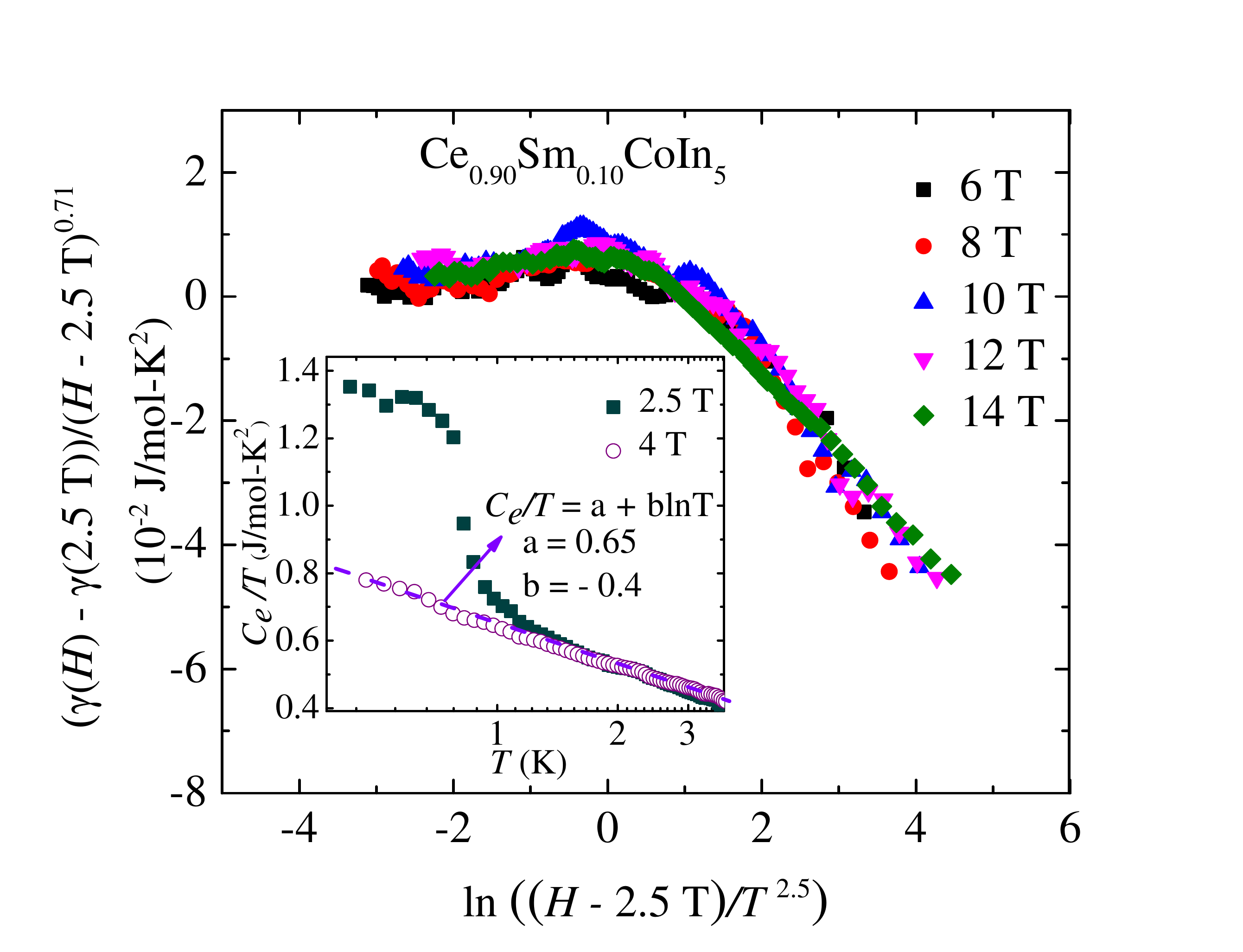}
\caption{(Color online) Scaling of the Sommerfeld coefficient $\gamma$ according to $\gamma(H)-\gamma(2.5$T$) \propto (H-2.5$T$)^{0.71} f [(H - H(2.5$T$)/T^{2.5}]$. We obtained the best scaling shown on the main panel with a logarithmic dependence of $\gamma(2.5$T) vs $T$ at temperatures $T\leq 6$ K. Inset: Semi-log plot of $\gamma \equiv C_e/T$ vs temperature $T$ measured at 2.5 T and 4 T and their normal state fit with $C_{e}/T$ = 0.65 - 0.4ln$T$.}
\label{C1F5}
\end{figure}

\paragraph{Quantum critical scaling of heat capacity.} When the system is tuned to a quantum critical point by a magnetic field $H=H_{\textrm{QCP}}$, the time-scale of the quantum critical fluctuations is governed by temperature only, i.e., $\tau=\hbar/k_BT$. As a consequence, the relevant dynamical response functions exhibit the 
$\omega/T$ scaling, where $\omega$ is the characteristic frequency on which the system is probed \cite{varma1989phenomenology,sachdev2007quantum}. When $H$ serves as a tuning parameter, the effect of quantum critical fluctuations in the thermodynamic or transport quantities is manifested in their dependence on the  ratio of $H-H_{\textrm{QCP}}$ and $T$, as well as a typical energy scale describing the source of quantum fluctuations. 

In order to further confirm that the anomalous temperature
dependence of $C_e/T$ is governed by quantum critical fluctuations and that $H_{QCP}= 2.5$ T for the $x=0.10$ samples, we show that $\gamma(H,T)\equiv C_e/T$ is governed by the critical free energy density $f_\textrm{cr}=a_0r^{\nu(d+z)}f_0(T/r^{\nu z})=a_0T^{(d+z)/z}\tilde{f}_0(r/T^{1/\nu z})$, where
$a_0$ is a constant, $f_0$ and $\tilde{f}_0$ are scaling functions, $r\propto (H-H_{\textrm{QCP}})$, $d$ is the dimensionality of the system, $z$ is the dynamical critical exponent, and $\nu$ is the critical exponent describing the dynamical correlations between Ce moments. Therefore, based on the arguments of Ref. \cite{tsvelik1993phenomenological} and the scaling analysis performed for CeCoIn$_5$  \cite{bianchi2003avoided}, we performed the scaling analysis  using  the function ($ \gamma(H)-\gamma(H_{\textrm{QCP}}))\propto (H-H_{\textrm{QCP}})^\alpha f[(H-H_{\textrm{QCP}})/T^\beta]$, where $\alpha \equiv \nu(d+z)$ and  $\beta$ represents the scaling dimension of $H$. The best scaling we obtained (Fig.~\ref{C1F5}) confirms that $H_{\textrm{QCP}} = 2.5$ T for Ce$_{0.90}$Sm$_{0.10}$CoIn$_5$ and gives $\alpha$ = 0.71, and $\beta$ = 2.5. The scaling of $\gamma(H,T)$ spans both the FL regime at low temperatures and the NFL regime  at high temperatures, with all five data sets for different $H$ values overlapping over the $T$ range 0.42 K $\leq T\leq$ 8 K. A very similar scaling has been observed in the stocheometric compound CeCoIn$_5$ \cite{bianchi2003avoided} and YbRh$_2$Si$_2$ \cite{trovarelli2000ybrh}

It is instructive to compare these results for the values of $\alpha$ and $\beta$ of Ce$_{0.90}$Sm$_{0.10}$CoIn$_5$ with those obtained for Ce$_{1-x}$Yb$_x$CoIn$_5$ \cite{singh2018zero}. For the latter, we found $\alpha_{\textrm{Yb}}=0.71$ and $\beta_{\textrm{Yb}}=1.2$, while for the former $\alpha_{\textrm{Sm}}=0.71$ and $\beta_{\textrm{Sm}}=2.5\approx 2\beta_{\textrm{Yb}}$. On the other hand, both $\alpha_{\textrm{Sm}}$ and $\beta_{\textrm{Sm}}$ match the corresponding values found for the stocheometric compound. Given the fact that magnetic field, on one hand, serves as a tuning parameter to the QCP, while, on the other hand, suppresses the magnetic fluctuations by direct coupling to the magnetic moments of Ce ions, the relatively high value of $\beta$ signals that the region of quantum critical fluctuations is broader for Ce$_{0.90}$Sm$_{0.10}$CoIn$_5$ compared to Ce$_{1-x}$Yb$_x$CoIn$_5$. This is also consistent with the stronger suppression of superconductivity in the samarium alloys.

\begin{figure}[!]
\includegraphics[width=3.3in,angle=0]{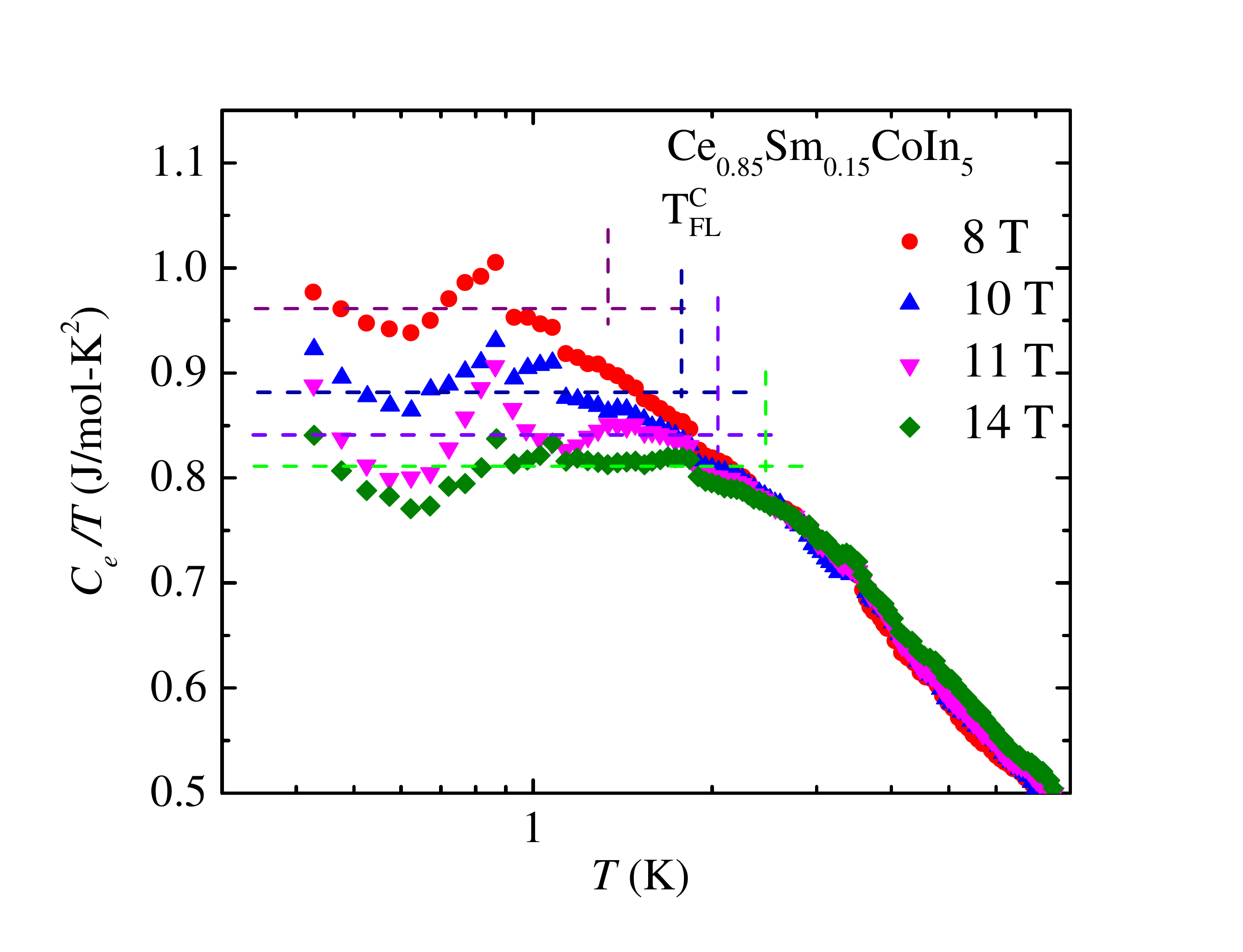}
\caption{(Color online) Semi-log plot of the electronic specific heat normalized by temperature $C_e/T$ vs temperature $T$ of Ce$_{0.85}$Sm$_{0.15}$CoIn$_5$ measured with applied magnetic fields $H\parallel c$-axis  over the temperature range 0.42 K $\leq T\leq$ 10 K. $T^C_{FL}$ represents the temperature at which the behavior changes from  Fermi liquid to non-Fermi liquid.}
\label{C1F6}
\end{figure}

A major obstacle in performing the scaling analysis was to determine the normal state $\gamma(T, 2.5$T$)$ at low temperatures because the $x$ = 0.10 sample exhibits  superconductivity below 1.1 K in the presence of a magnetic field of 2.5 T. We were able to overcome this problem by determining  the $T$ dependence of the $\gamma(T,4$T) down to 0.42 K and taking advantage of the fact that $\gamma(T,4$T) and $\gamma(T,2.5$T) completely overlap in the normal state, i.e., for $T>1.2$ K (see inset to Fig~\ref{C1F5}). 
We found that the $C_e(T, H)/T$ follows a logarithmic $T$-dependence with $C_e(T, H)/T$  = 0.64 + ln$T^{-0.4}$. This result implies that proximity to the underlying quantum critical point affects the thermodynamic properties significantly. 

\subsection{Ce$_{0.85}$Sm$_{0.15}$CoIn$_5$}
\paragraph{Heat capacity.} In order to determine the effect of doping on the value of $H_{\textrm{QCP}}$ and to search for a critical Sm concentration for which $H_{QCP}=0$, we performed similar specific heat and resistivity measurements as a function of temperature and magnetic field on single crystals with a slightly higher Sm concentration, i.e. Ce$_{0.85}$Sm$_{0.15}$CoIn$_5$. Figure~\ref{C1F6} shows the temperature dependence of $C_e/T$  measured in fields 8 T $\leq H\leq$ 14 T and at low temperatures, i.e., 0.4 K $\leq T\leq$ 8 K. The normal-state results shown in Fig.~\ref{C1F6} reveal sharp  crossovers from constant $C_e/T$ (FL at low $T$) to logarithmic $T$-dependent $C_e/T$, with $C_e/T=1 + ln T^{-0.64}$ ( NFL at high $T$). The crossovers temperature $T^{\mathrm{C}}_{\mathrm{FL}}$ also  shifts towards higher temperatures with increasing $H$ for $H\geq$ 6 T.

  \begin{figure}[!]
\includegraphics[width=3.3in,angle=0]{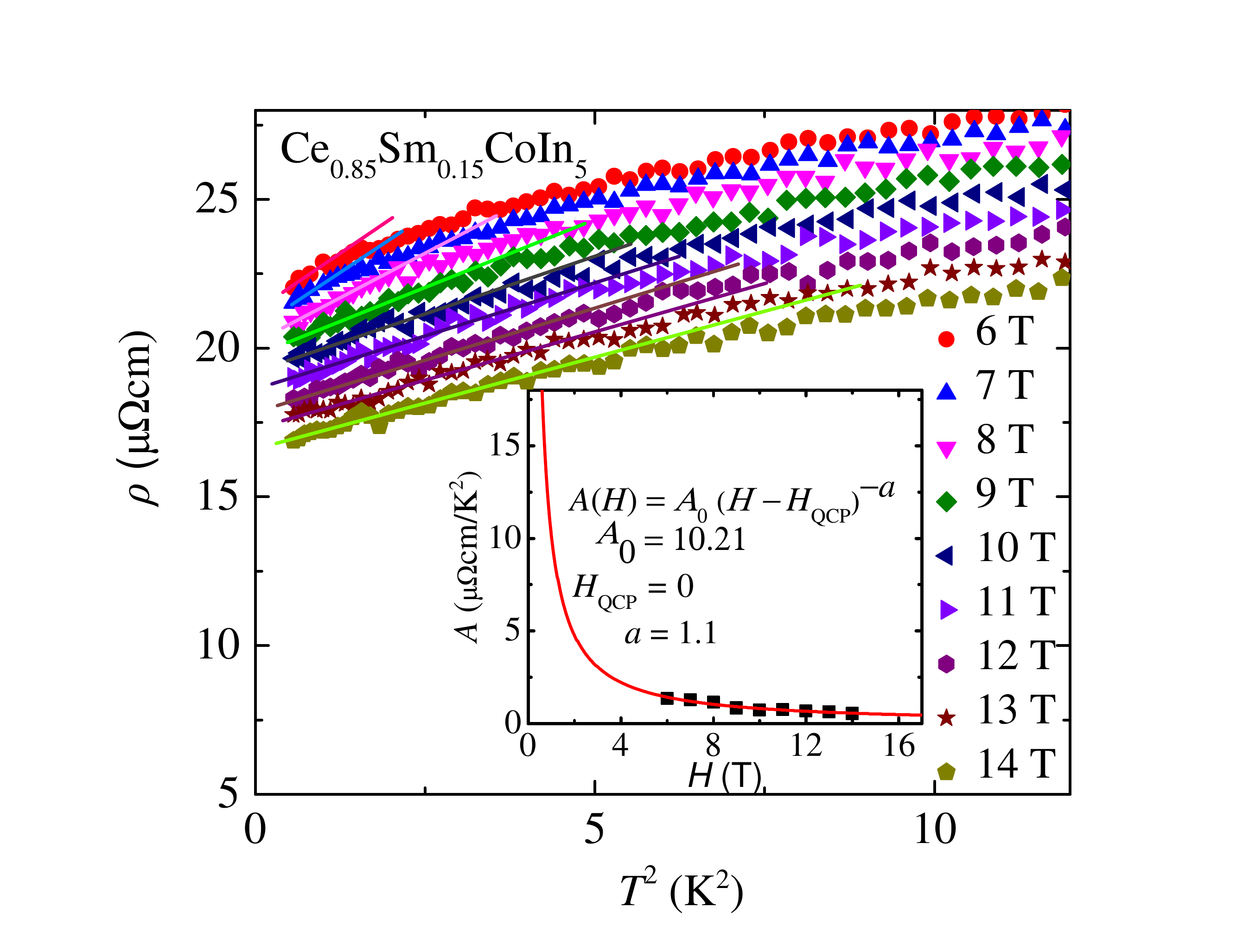}
\caption{(Color online) Low temperature resistivity $\rho$ vs temperature $T^2$ of Ce$_{0.85}$Sm$_{0.15}$CoIn$_5$ measured in a magnetic field $H$ with 6 T $\leq H\leq$ 14 T. The solid lines are linear fits to the data with $\rho(T,H) = \rho_0 + AT^2$. Inset: $H$ dependent coefficient $A$ of $\rho(T,H)$.}
\label{C1F7}
\end{figure}

\paragraph{Resistivity.} Figure~\ref{C1F7} shows the $T^2$ dependence of $\rho(H,T)$ of Ce$_{0.85}$Sm$_{0.15}$CoIn$_5$ measured from 0.6 K to 5 K and for  6 T $\leq H \leq$ 14 T. As in the case of the single crystals with $x$ = 0.10, the resistivity for the $x=0.15$ samples is linear in $T^2$ at low temperature, signaling the presence of the Fermi-liquid behavior. The $H$ dependence of the coefficient $A$ (see inset to Fig.~\ref{C1F7}) is  best fitted with the  function  $A(H) \propto(H-H_{\textrm{QCP}})^{-a}$, with $H_{QCP}$ = 0 and  $a$ = 1.1. Since $A(H)$ vs $H$ curve diverges for $H_{QCP}$ = 0,   Ce$_{0.85}$Sm$_{0.15}$CoIn$_5$ is quantum critical.
    
\begin{figure}[!]
\includegraphics[width=3.3in,angle=0]{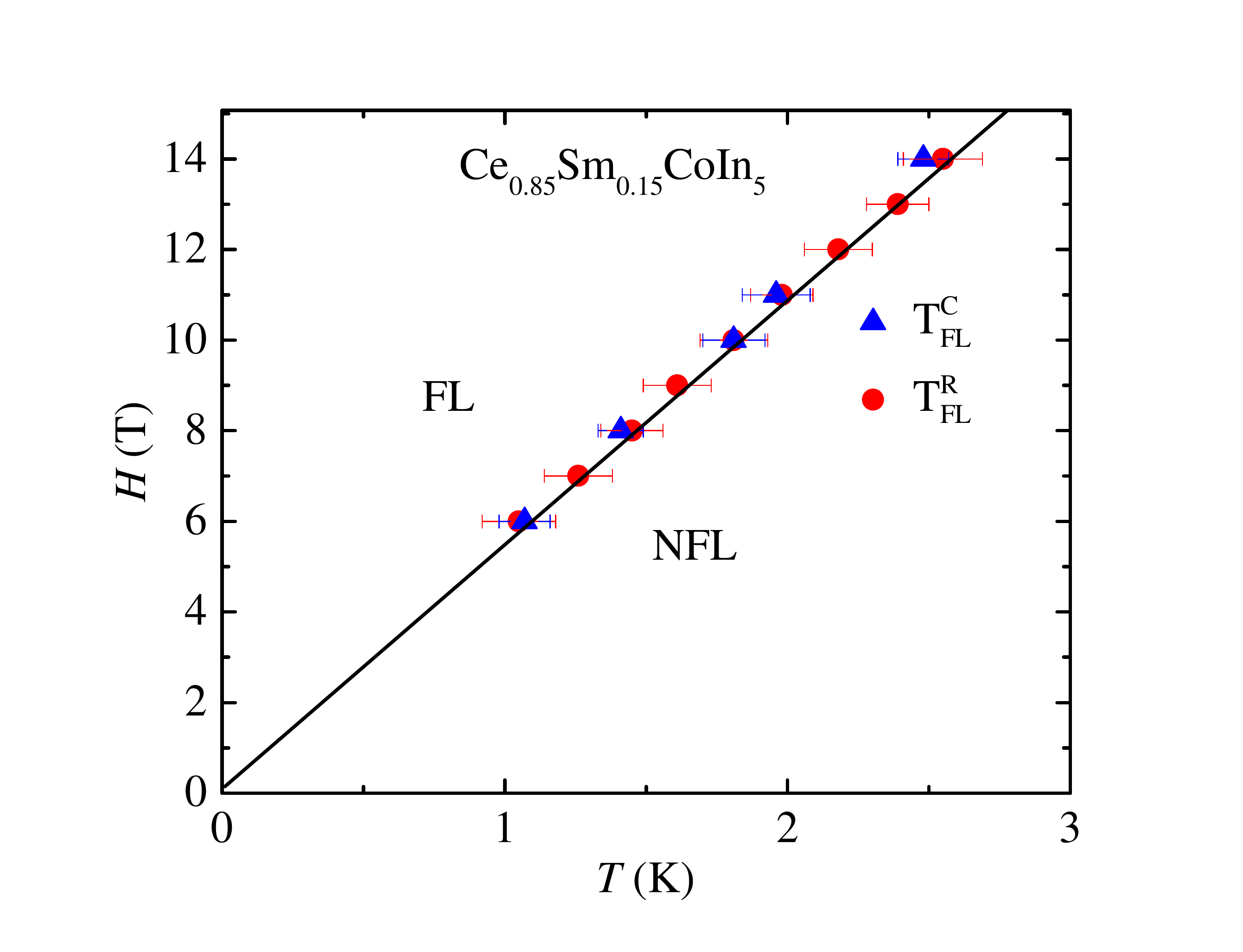}
\caption{(Color online) Magnetic field-temperature $(H-T)$ phase diagram of Ce$_{0.85}$Sm$_{0.15}$CoIn$_5$ with $H\perp ab$ plane.}
\label{C1F8}
\end{figure}
\begin{figure}[!]
\begin{center}
\includegraphics[width=3.3in,angle=0]{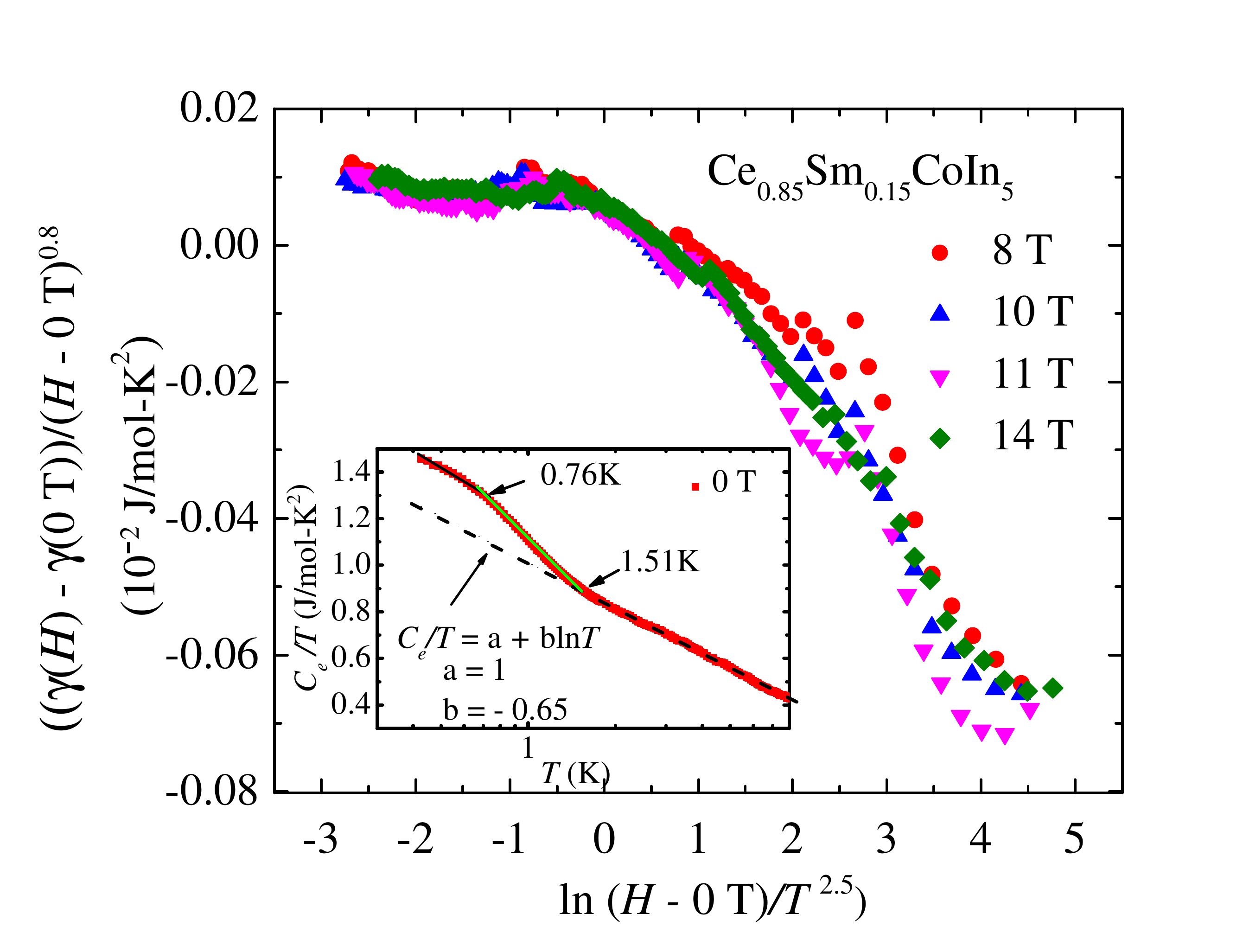}
\caption{(Color online) Scaling of the Sommerfeld coefficient $\gamma$ according to $\gamma(H) - \gamma(0$T$) \propto H^{0.8} f (H)/T^{2.5}$ with $\gamma$ = $C_e/T$. We obtained the best scaling shown on the main panel with a logarithmic dependence of $\gamma(0$T) vs $T$ for temperatures $T\leq$ 6 K. Inset: Semi-log plot of $\gamma \equiv C_e/T$ vs temperature $T$ measured at 0 T and its normal state fit with $C_e(T,0)/T$ = 1 - 0.65 ln$T$.}
\label{C1F9}
\end{center}
\end{figure}

To check that the samples with $x=0.15$ are, indeed, quantum critical, we plotted in Fig.~\ref{C1F8} the $H-T$ phase diagram for this Sm-doped single crystals. The FL to NFL crossovers extracted from $C_e/T$ are represented by $T^{\mathrm{C}}_{\mathrm{FL}}$ (solid blue triangles), and these crossovers extracted from $\rho(T,H)$ are represented by $T^{\mathrm{R}}_{\mathrm{FL}}$(solid red circles). The  FL to NFL crossovers  obtained from both measurements are in excellent agreement. The linear extrapolation of the fit of these crossovers to zero temperature reveals that $H_{QCP}$ = 0, indicating that the sample Ce$_{0.85}$Sm$_{0.15}$CoIn$_5$ is, indeed, at the critical doping.

\paragraph{Quantum critical scaling of heat capacity.} We further checked whether Ce$_{0.85}$Sm$_{0.15}$CoIn$_5$ is quantum critical by performing the scaling analysis as discussed above. We obtained the best scaling, shown in Fig.~\ref{C1F9}, with $H_{\textrm{QCP}} = 0$, $\alpha = 0.8$ and $\beta = 2.5$. The scaling of $\gamma \equiv C_e/T$ spans both the FL at low temperatures and the NFL regime  at high temperatures, with all four data sets measured at different $H$ values overlapping over the temperature range 0.42 K $\leq T\leq5$ K. This scaling further indicates  that Ce$_{0.85}$Sm$_{0.15}$CoIn$_5$ is quantum critical and exhibits NFL behavior.

 In order to determine $\gamma(0$T) required for the scaling of Fig.~\ref{C1F9}, we fitted the $C_e(T$,0)$/T$ data for $T>1.91$ K of the $x$ = 0.15 sample with a logarithmic $T$-dependence; i.e., $C_e(T$,0)$/T$ = 1 + ln$T^{-0.64}$. The data for $T < 1.5$ K display a stronger than logarithmic increase with decreasing $T$, most likely due to the presence of long range AFM fluctuations and superconductivity. This is consistent with the phase diagram of Fig.~\ref{C1F10}, where $T_N=0.8$ K. Nevertheless, we used the logarithmic $T$ dependence of $\gamma(0$T) obtained by fitting the data at $T>1.91$ K over the whole $T$ range down to 0.42 K 

Combining the experimental results shown above, we have generated an  $H_{\textrm{QCP}}$ vs Sm concentration  phase diagram of  Ce$_{1-x}$Sm$_{x}$CoIn$_5$, for the doping $x$ = 0.00, 0.10, 0.15, as shown in the inset to Fig.~\ref{C1F10}. This phase diagram shows that $H_{\textrm{QCP}}$ is suppressed with increasing Sm$^{3+}$ doping and becomes zero for $x\geq$ 0.15. Therefore, at zero temperature in the region for $x < 0.1$ (main panel of Fig.~\ref{C1F10}) superconductivity and long-range AFM order of Ce moments coexist. For $0.1\leq x\leq$ 0.15, superconductivity, long-range AFM order of Ce moments, and long-range AFM of Sm ions coexist at zero temperature. In region $0.15\leq x\leq$ 0.17 superconductivity co-exists with long-range AFM order of Sm moments, while there is only long range AFM order due to  Sm ions when $x\geq 0.17$. 
 
\section{Discussion and outlook}
Anomalous thermodynamic and transport low-temperature properties of complex materials have long been associated with underlying QCPs. In this regard, CeCoIn$_5$ as well as the other members of the '115' family of compounds are not exception. The existence of the $H_{\textrm{QCP}}$ in CeCoIn$_5$ has been already established independently by several groups. Our present study further justifies the validity of using the conceptual framework of quantum criticality to account for the observed anomalies in Ce$_{1-x}$Sm$_x$CoIn$_5$. 

\begin{figure}[h]
\centering
\includegraphics[width=0.9\linewidth]{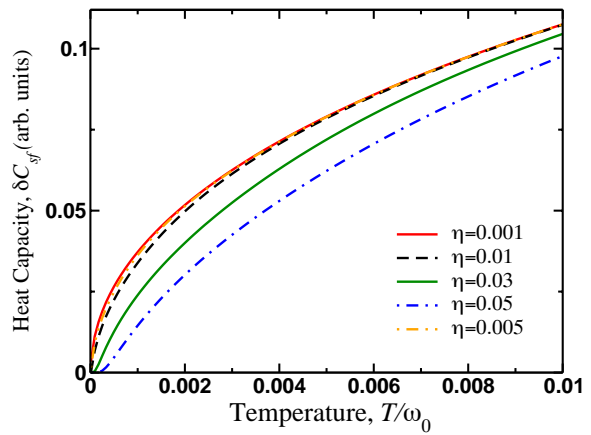}
\caption{(Color online) Fluctuation correction to the heat capacity due to system's proximity to magnetic QCP as a function of temperature for various values of the parameter $\eta\propto (H-H_{\textrm{QCP}})$. The temperature is given in the units of the characteristic energy scale for the magnetic subsystem. Although $\delta C_{\textrm{sf}}(T)\propto T^\alpha$ has a power-law (and not logarithmic) temperature dependence, the exponent $\alpha$ shows strong dependence on the parameter $\eta$, a feature which is also observed experimentally. }
\label{FigdCsf}
\end{figure}

We can also estimate the fluctuation correction to the heat capacity in external $H$. The details of the calculation are given in the Appendix A, so here we only present the results in Fig. \ref{FigdCsf}.  We found that due to presumably strong coupling between itinerant carriers and localized moments of Ce ions, the fluctuation correction to the heat capacity in the close vicinity to QCP does have a power-law temperature dependence $\delta C_{\textrm{sf}}\propto T^{\alpha}$ with $\alpha \approx 0.45$. When the system is de-tuned from the proximity to the QCP, the exponent $\alpha$ increases and becomes $\alpha\approx 1.5$ at very low temperatures. 

Samarium substitution on cerium sites brings about a novel feature into the phase diagram: the possible co-existence between superconductivity and antiferromagnetic ordering of Sm local moments in the narrow region of 
$0.1\leq x\leq 0.17$, Fig 1. The N\'{e}el temperature vanishes at $x\approx 0.1$ giving rise most likely to an antiferromagnetic QCP. In this regard we would like to note that quantum critical fluctuations may already manifest themselves in the superconducting state.  Indeed, recent theoretical works have shown the the effect of quantum critical fluctuations can be also probed in the superconducting state by studying the temperature dependence of thermodynamic functions such as heat capacity \cite{Fernandes2020,Khodas2020} and London penetration depth \cite{Khodas2020}. In particular, it has been shown that in a fully gapped superconductor, quantum fluctuations produce power-law dependences in both of these quantities on the background of the (mean-field) exponential temperature dependence. In the context of Ce$_{1-x}$Sm$_x$CoIn$_5$, it would be intriguing to probe the variation in the temperature dependence of the heat capacity of the $x=0.1$ samples in the superconducting state. From the point of view of the theoretical analysis, this problem poses an additional challenge since the fermionic spectrum is not fully gaped.

\section{Summary}
Through the measurements of specific heat, resistivity, and magnetoresistivity on both Ce$_{0.90}$Sm$_{0.10}$CoIn$_5$ and Ce$_{0.85}$Sm$_{0.15}$CoIn$_5$ samples, we observe that the FL regime recovery is established  with increasing $H$ in both $\rho(H,T)$ and $C_e(H,T)/T$ along with the low T evolution of positive magnetoresistivity. We conclude that $H_{\textrm{QCP}}$ decreases with increasing Sm concentration. The single crystal with $x$ = 0.15 exhibits zero temperature quantum criticality associated with the antiferromagentic ground state of Ce ions. However, the lower doped crystal Ce$_{0.90}$Sm$_{0.10}$CoIn$_5$ reveals $H_{\textrm{QCP}}$ at 2.5 T. In addition, crossovers ($T\approx19$ K) from positive to negative MR extending to higher temperatures are the evidence that the change of spin fluctuations with $H$ may be strongly associated to the critical nature of the slope $A(H)$ of 
$\rho(H,T)$, suggesting that the QCP is magnetic in nature. As compared to the parent compound and low doped sample ($x$ = 0.1), the normal-state transport and thermodynamic properties of Ce$_{0.85}$Sm$_{0.15}$CoIn$_{5}$ are controlled by the presence of QCP alone. Moreover, the scaling analysis of the $C_e/T$ data provides strong evidence for the existence of $H_{\textrm{QCP}}$. Similarly, excellent fits of  $C_e(H,T)/T$ and $\rho(T,H)$ data measured at several $H$ also suggest that the QCP is antiferromagnetic in nature as supported by the spin fluctuation theory.

\section{Acknowledgments} 
The work was supported by the National Science Foundation under grants Nos. DMR-1904315 and NSF-DMR-BSF-2002795 at Kent State University, and by the US Department of Energy, Office of Basic Energy Sciences, Division of Materials Sciences and Engineering, under Grant No. DE-FG02-04ER46105 at UCSD.

\begin{appendix}
\section{Temperature correction to the heat capacity from quantum critical spin-fluctuations}
The contribution of the quantum critical spin-fluctuations to the free energy is \cite{Fernandes2020,Khodas2020}: 
\begin{equation}\label{Eq1}
\delta F_{\textrm{sf}}=3T\sum\limits_{m=-\infty}^{\infty}\int\frac{d^2{\mathbf q}}{(2\pi)^2}\log\left[\chi^{-1}({\mathbf q},\Omega_m)\right].
\end{equation}
Here $\chi^{-1}({\mathbf q},\Omega_m)=\nu\left(E_\bq^2-i|\Omega_m|/\omega_{\textrm{sf}}+\Omega_m^2/\omega_0^2\right)$, $E_\bq=\sqrt{\eta+({\mathbf q}/Q_0)^2}$, $\nu$ is inverse proportional to the static spin-susceptibility at $T=0$, parameter $\eta$ controls the proximity to the magnetic QCP, $\Omega_m=2\pi Tm$ is the bosonic Matsubara frequency, $\omega_{\textrm{sf}}$ is an energy scale which describes the interaction between the localized moments and conduction conductions, $\omega_0$ is a typical energy scale of the magnetic system \cite{Abanov2000,Metlitski2010} and the numerical pre-factor takes into account three fluctuating modes: two transverse and one longitudinal. 

Since the spin-fluctuation propagator is a non-analytic function of $\Omega_m$, in order to evaluate the free energy we will need to keep the real part of the expression under the integrals only. 
In the Matsubara summation (\ref{Eq1}) we can single out the $m=0$ term and reduce the remaining summation over $m\geq 1$:
\begin{widetext}
\beg\label{AEq1}
\delta F_{\textrm{sf}}=3T\int\frac{d^2\bq}{(2\pi)^2}\log(\nu E_\bq^2)+6T\int\frac{d^2\bq}{(2\pi)^2}\log\left[
\prod\limits_{n=1}^\infty\left(1+\frac{\omega_0^2E_\bq^2}{\Omega_n^2}\right)\prod\limits_{m=1}^\infty
\left(\frac{\nu\Omega_m^2}{\omega_0^2}\right)\prod\limits_{l=1}^\infty
\left(1-\frac{i\Omega_l/\omega_{\textrm{sf}}}{E_\bq^2+\frac{\Omega_l^2}{\omega_0^2}}\right)\right].
\en
\end{widetext}
The first term in this expression represents a "zero-point-motion" correction and, therefore, does not produce the temperature dependent contribution to the heat capacity.

In the second term, there are three products under the logarithm which we will discuss them separately below.  The first product evaluates to 
\beg\label{AEq2}
\begin{split}
&\prod\limits_{m=1}^\infty\left(1+\frac{\omega_0^2E_\bq^2}{\Omega_m^2}\right)=\frac{\sinh(\pi z)}{\pi z}, 
\quad z=\frac{\omega_0E_\bq}{2\pi T}.
\end{split}
\en
To evaluate the second product we need to use regularization scheme to assign it a finite value. We use the zeta-function regularization scheme \cite{Khodas2020} to find
\beg\label{AEq3}
\sum\limits_{m=1}^\infty
\log\left(\frac{\nu\Omega_m^2}{\omega_0^2}\right)=\log\left(\frac{\omega_0}{\sqrt{\nu}T}\right).
\en
Finally, for the third product we clearly need to evaluate the real part only. It obtains: 
\beg\label{AEq4}
\begin{split}
\prod\limits_{l=1}^\infty\sqrt{1+\frac{(\Omega_l/\omega_{\textrm{sf}})^2}{\left(E_\bq^2+\frac{\Omega_l^2}{\omega_0^2}\right)^2}}=\frac{\sinh(r_\bq^{+})\sinh(r_\bq^{-})}{2\sinh^2(\pi z)},
\end{split}
\en
where we introduced auxiliary variables 
\beg\label{AEq5}
r_\bq^{\pm}=\frac{\omega_0^2}{2\sqrt{2}\omega_{\textrm{sf}}T}\sqrt{1+\frac{2\omega_{\textrm{sf}}^2}{\omega_0^2}E_\bq^2\pm\sqrt{1+\frac{4
\omega_{\textrm{sf}}^2}{\omega_0^2}E_\bq^2}}.
\en
Thus, the expression for the fluctuation correction to the free energy is
\beg\label{AEq6}
\delta F_{\textrm{sf}}=6T\int\frac{d^2\bq}{(2\pi)^2}\log\left[\frac{\sinh(r_\bq^{+})\sinh(r_\bq^{-})}{\sinh(\omega_0E_\bq/2T)}\right].
\en
The fluctuation correction to the heat capacity is directly obtained from (\ref{AEq6}), $\delta C_{\textrm{sf}}/T=-\partial^2(\delta F_{\textrm{sf}})/\partial T^2$. We find
\beg\label{AEq7}
\begin{split}
\delta C_{\textrm{sf}}(T)=6\int\frac{d^2\bq}{(2\pi)^2}&\left[\frac{(r_{\bq}^{+})^2}{\sinh^2(r_\bq^{+})}+\frac{(r_{\bq}^{-})^2}{\sinh^2(r_\bq^{-})}\right.\\&\left.-\frac{\omega_0^2E_{\bq}^2}{4T^2\sinh^2(\omega_0E_\bq/2T)}\right].
\end{split}
\en
Our numerical analysis of this expression shows that at low temperatures, $T\ll \omega_0$ and $\omega_{\textrm{sf}}\approx \omega_0$, $\delta C_{\textrm{sf}}(T)\propto T^\alpha$ with $\alpha\approx 0.45$ at the QCP.

\end{appendix}

\bibliography{Ref_skutt}

\end{document}